\documentclass[a4paper]{article}
\usepackage[margin=2.54cm]{geometry}
\usepackage[T1]{fontenc}
\usepackage{lmodern}
\usepackage{amsmath}
\usepackage{amssymb}
\usepackage{abstract}

\usepackage[linktocpage,pdftitle={Thermal Inflation with a Thermal Waterfall Scalar Field Coupled to a Light Spectator Scalar Field},pdfauthor={Konstantinos Dimopoulos, David H. Lyth and Arron Rumsey}]{hyperref}
\usepackage[dvipsnames]{xcolor}
\definecolor{hyperlinks}{HTML}{0000FF}
\hypersetup{colorlinks, urlcolor=hyperlinks, citecolor=hyperlinks, linkcolor=hyperlinks, breaklinks=true}
\usepackage{breakurl}
\usepackage{graphicx}
\graphicspath{ {Images/} }

\usepackage{cite}

\usepackage{enumitem}
\setlist{listparindent=\parindent}
\usepackage{cleveref}

\crefname{chapter}{Chapter}{Chapters}
\Crefname{chapter}{Chapter}{Chapters}
\crefname{section}{Section}{Sections}
\Crefname{section}{Section}{Sections}
\crefname{appendix}{Appendix}{Appendices}
\Crefname{appendix}{Appendix}{Appendices}
\crefname{equation}{Eq.}{Eqs.}
\Crefname{equation}{Eq.}{Eqs.}
\crefname{figure}{Fig.}{Figs.}
\Crefname{figure}{Fig.}{Figs.}
\crefname{table}{Table}{Tables}
\Crefname{table}{Table}{Tables}
\numberwithin{equation}{section}
\usepackage[title,titletoc]{appendix}
\title{\bf Thermal Inflation with a Thermal Waterfall Scalar Field Coupled to a Light Spectator Scalar Field}
\author
{
Konstantinos Dimopoulos$^{a}$, David H. Lyth$^{b}$ and Arron Rumsey$^{c}$
\\
\\
{\small\em Consortium for Fundamental Physics}
\\
{\small\em Physics Department, Lancaster University}
\\
{\small\em Lancaster, LA1 4YB, UK}
\\
\\
\small $^a$\ \href{mailto:konst.dimopoulos@lancaster.ac.uk}{konst.dimopoulos@lancaster.ac.uk}, $^b$\ \href{mailto:d.lyth@lancaster.ac.uk}{d.lyth@lancaster.ac.uk}, $^c$\ \href{mailto:a.rumsey@lancaster.ac.uk}{a.rumsey@lancaster.ac.uk}
}
\date{\today}
\begin{document}
\maketitle
\vspace{-1.5em}
\begin{abstract}
\noindent
A new model of thermal inflation is introduced, in which the mass of the 
thermal waterfall field is dependent on a light spectator scalar field. 
Using the $\delta N$ formalism, the ``end of inflation'' scenario is 
investigated in order to ascertain whether this model is able to produce the 
dominant contribution to the primordial curvature perturbation. 
A multitude of constraints are considered so as to explore the parameter space, 
with particular emphasis on key observational signatures. For natural values of
the parameters, the model is found to yield a sharp prediction for the scalar 
spectral index and its running, well within the current observational bounds.
\end{abstract}

\begin{twocolumn}

\section{Introduction}
\label{Section: Introduction}
Cosmological Inflation is the leading candidate for the solution of the three 
main problems of the standard Big Bang cosmology: the horizon, flatness and 
relic problems. It also has the ability to seed the initial conditions required
to explain the observed large-scale structure of the Universe
\cite{Lyth_&_Liddle:Prim._Den._Pert.}.
In the simplest scenario, quantum fluctuations of a scalar field are converted 
to classical perturbations around the time of horizon exit, after which they 
become frozen. This gives rise to the primordial curvature perturbation, 
$\zeta$, which grows under the influence of gravity to give rise to the 
large-scale structure in the Universe. 
The simple single-field inflationary scenario is favoured by current 
observations \cite{Planck_2015_Cosmo._Results}. 
However, given the richness and complexity of the theories 
beyond the standard model, this simple picture seems unlikely.

Moving away from this simplest scenario, there has been much work done on
generating the observed $\zeta$ in other scenarios, such as the curvaton 
\cite{Lyth_&_Wands:Generating_Curv._Pert._without_Inflaton , 
Lyth_et_al.:Prim._Den._Pert._in_Curvaton_Scenario , 
Choi_&_Seto:Modulated_Reheat._Curvaton , 
Dimopoulos:Can_Vec._Field_be_respons._Cur._Pert._Uni. , 
Dimopoulos_et_al.:Stat._Ani._Curv._Pert._Vec._Field_Perts. , 
Dimopoulos:Stat._Ani._Vec._Curv._Paradigm , 
Yokoyama_&_Soda:Prim._Stat._Ani._generated_End_of_Inf. , 
Assadullahi_et_al.:Mod._Curvaton_Dec. , 
Langlois_&_Takahashi:Den._Perts._Mod._Dec._Curvaton , 
Enomoto_et_al.:Mod._Dec._Multi-Comp._Uni. , 
Kohri_et_al.:Delta-N_Form._Curvaton_Mod._Decay , 
Dimopoulos_et_al.:Curvaton_Dyn.}, 
inhomogeneous reheating \cite{%
Dvali_et_al.:New_Mech._Generating_Den._Perts._from_Inflation , 
Dvali_et_al.:Cosmo._Perts._Inhomo._Reheat._Freezeout_&_Mass_Dom. , 
Postma:Inhom._Reheat._Low_Scale_Inflation_and/or_MSSM_Flat_Dirs. , 
Choi_&_Seto:Modulated_Reheat._Curvaton , 
Vernizzi:Generating_Cosmo._Perts._Mass_Variations , 
Vernizzi:Cosmo._Perts._Varying_Masses_&_Couplings , 
Kawasaki_et_al.:Den._Fluctuations_in_Thermal_Inflation_&_Non-Gaussianity , 
Assadullahi_et_al.:Mod._Curvaton_Dec. , 
Langlois_&_Takahashi:Den._Perts._Mod._Dec._Curvaton , 
Enomoto_et_al.:Mod._Dec._Multi-Comp._Uni. , 
Kohri_et_al.:Delta-N_Form._Curvaton_Mod._Decay}, ``end of inflation'' 
\cite{Lyth:Generating_Curvature_Perturbation_E._of_I. , 
Yokoyama_&_Soda:Prim._Stat._Ani._generated_End_of_Inf. , 
Kawasaki_et_al.:Den._Fluctuations_in_Thermal_Inflation_&_Non-Gaussianity , 
Salem:On_Gen._Den._Perts._End_of_Inf. , 
Alabidi_&_Lyth:Curv._Pert._Sym._Break._End_of_Inf. , 
Lyth:Hybrid_Waterfall_and_Curvature_Perturbation , 
Lyth_&_Riotto:Gen._Curv._Pert._End_of_Inf._Str._Theory , 
Sasaki:Multi-brid_Inf._Non-Gauss.} 
(also see \cite{Bernardeau_et_al.:Mod._Flucts._Hyb._Inf.}) and inhomogeneous 
phase transition 
\cite{Matsuda:Cosmo._Perts._from_Inhomogeneous_Phase_Transition} (also see 
\cite{%
Alabidi_et_al.:How_Curvaton_Mod._Reheat._and_Inhomo._End_of_Inf._are_related}).

One particular model of inflation is thermal inflation 
\cite{Lyth_&_Stewart:Cos._TeV_Mass_Higgs_Field_break._GUT_Gauge._Sym. , 
Lyth_&_Stewart:Therm._Inf._Moduli_Prob. , 
Barreiro_et_al.:Aspects_Therm._Inf.:_Finite_Temp._Pot._Top._Defects , 
Asaka_&_Kawasaki:Cos._Moduli_Problem_Therm._Inf._Models}, which is a brief 
period of inflation 
that could have occurred after a period of prior primordial inflation. 
Thermal inflation lasts too little to solve the problems of the standard Big 
Bang cosmology that motivate primordial inflation, but it may be rather useful 
to dilute any dangerous relics that are not dealt with by primordial inflation 
such as moduli fields or gravitinos. Another interesting byproduct of thermal 
inflation is changing the number of e-folds before the end of primordial 
inflation, which correspond to the cosmological scales. This has an immediate 
effect on inflationary observables and can assist in inflation model building
\cite{charlotte,charlotte+}.

Thermal inflation occurs due to finite-temperature effects arising from a 
coupling between a so-called thermal waterfall scalar field $\phi$ and the 
thermal bath created from the partial or complete reheating from primordial 
inflation. Thermal field theory gives a thermal contribution 
$g^{2}T^{2}\phi^{2}$ to the effective scalar potential, 
where $g$ is the coupling constant of the interaction between 
$\phi$ and the thermal bath and $T$ is the bath's temperature.
This results in a thermal correction to the effective mass of $g^{2}T^{2}$.
This thermal mass can temporarily trap the thermal waterfall field on top of
a false vacuum, resulting in inflation. However, as time goes by, the thermal 
mass of $\phi$ decreases such that a phase transition sends $\phi$ to its 
vacuum expectation value (VEV) and inflation is terminated.

Despite occurring much later than primordial inflation, thermal inflation may
produce a substantial contribution to the curvature perturbation. This is how. 
The mass of a given scalar field may depend on the expectation value of another 
scalar field.
\cite{Vernizzi:Generating_Cosmo._Perts._Mass_Variations , 
Vernizzi:Cosmo._Perts._Varying_Masses_&_Couplings , 
Lyth:Generating_Curvature_Perturbation_E._of_I. , 
Matsuda:Cosmo._Perts._from_Inhomogeneous_Phase_Transition , 
Yokoyama_&_Soda:Prim._Stat._Ani._generated_End_of_Inf. , 
Dvali_et_al.:Cosmo._Perts._Inhomo._Reheat._Freezeout_&_Mass_Dom. , 
Postma:Inhom._Reheat._Low_Scale_Inflation_and/or_MSSM_Flat_Dirs. , 
Salem:On_Gen._Den._Perts._End_of_Inf. , 
Alabidi_&_Lyth:Curv._Pert._Sym._Break._End_of_Inf. , 
Lyth:Hybrid_Waterfall_and_Curvature_Perturbation , 
Alabidi_et_al.:How_Curvaton_Mod._Reheat._and_Inhomo._End_of_Inf._are_related , 
Kohri_et_al.:Delta-N_Form._Curvaton_Mod._Decay}. 
More specifically, the mass of a thermal waterfall field $\phi$ that is 
responsible for a bout of thermal inflation could be dependent on another 
scalar field $\psi$. We will call this $\psi$ a spectator field, because it 
needs not affect the dynamics of the Universe at any time. If $\psi$ is light 
during primordial inflation, its quantum fluctuations are converted to almost 
scale-invariant classical field perturbations at around the time of horizon 
exit. If $\psi$ remains light all the way up to the end of thermal inflation, 
then thermal inflation will end at different times in different parts of the 
Universe, because the value of the spectator field determines the mass of the 
thermal waterfall field $\phi$, which in turn determines the end of thermal 
inflation. This is the ``end of inflation'' mechanism 
\cite{Lyth:Generating_Curvature_Perturbation_E._of_I.} and it can generate a 
contribution to the primordial curvature perturbation $\zeta$. The motivation 
of this work is to explore this scenario to see if it can produce the dominant 
contribution to the primordial curvature perturbation with characteristic 
observational signatures, in which the inflaton's contribution to the 
perturbation can be ignored.\footnote{This paper is based on the original 
research that was conducted as part of the thesis \cite{My_Thesis}. This 
research has not been published elsewhere.}
As such, inflation model building is liberated
from the requirements to generate $\zeta$, which substantially reduces 
fine-tuning and renders viable many otherwise observationally excluded inflation
models \cite{liber}.

It should be noted that this scenario is very similar to that in 
Ref.~\cite{%
Kawasaki_et_al.:Den._Fluctuations_in_Thermal_Inflation_&_Non-Gaussianity}. 
However, in that paper the authors use a modulated coupling constant 
rather than a modulated mass. Also, the treatment that has been given to the 
work in this paper is much more comprehensive. One example of this is in the 
consideration of the effect that the thermal fluctuation of the thermal 
waterfall field has on the model 
(see \cref{Subsubsection: Thermal Fluctuation of phi}). Another example is the 
requirement that the thermal waterfall field is thermalized 
(see \cref{Subsubsection: Thermalization of phi}). Also, there is no 
consideration given in Ref.~\cite{%
Kawasaki_et_al.:Den._Fluctuations_in_Thermal_Inflation_&_Non-Gaussianity} 
to requiring a fast transition from thermal inflation to thermal waterfall field
oscillation (see 
\cref{Subsubsection: Time of Transition from Thermal Inflation to Thermal 
Waterfall Field Oscillation}), as detailed in 
Ref.~\cite{Lyth:Hybrid_Waterfall_and_Curvature_Perturbation}, 
as this paper appeared after 
Ref.~\cite{%
Kawasaki_et_al.:Den._Fluctuations_in_Thermal_Inflation_&_Non-Gaussianity}. 

This paper is structured as follows. In 
\cref{Section: A New Thermal Inflation Model} we introduce our new model. 
In \cref{Section: phi Decay Rate Spectral Index and Tensor Fraction} we give 
expressions for key observational quantities that are predicted by the model. 
In \cref{Section: ``End of Inflation'' Mechanism} we explore the 
``end of inflation'' scenario and obtain in detail a multitude of constraints 
for our model parameters. We conclude in \cref{Section: Conclusions}.

Throughout this work, natural units are used where 
$c\!=\!\hbar\!=\!k_{B}\!=\!1$ and 
Newton's gravitational constant is \mbox{$8\pi G=M_P^{-2}$}, with
$M_{P}\!=\!2.436\times10^{18}\ \mbox{GeV}$ being the reduced Planck Mass.

\section{A new Thermal Inflation model}
\label{Section: A New Thermal Inflation Model}

The potential that we consider in our model is
\begin{align}
V(\phi,\psi,T) = & V_{0}+ 
\left(g^2T^2-\frac12m_0^2+
h^{2}\frac{\psi^{2\alpha}}{M_{P}^{2\alpha-2}}\right)\phi^{2}\nonumber\\
& +\lambda\frac{\phi^{2n+4}}{M_{P}^{2n}} + \frac{1}{2}m_{\psi}^{2}\psi^{2},
\label{Eq.: Full Potential0}
\end{align}
where $\phi$ is the thermal waterfall scalar field, $\psi$ is a light spectator
scalar field, $T$ is the temperature of the thermal bath, $g$, $h$ and $\lambda$
are dimensionless coupling constants, $\alpha\ge1$ and $n\ge1$ are integers, 
$V_0$ is a density scale (corresponding to the scale of thermal inflation) and
 the $-m_{0}^{2}$ and $m_{\psi}^{2}$ are soft mass-squared terms coming form 
supersymmetry (SUSY) breaking.
A $\phi^4$ term is not featured because the thermal waterfall field is a flaton,
whose potential is stabilised by the higher-order non-renormalisable term
\cite{Lyth_&_Stewart:Cos._TeV_Mass_Higgs_Field_break._GUT_Gauge._Sym.,
Lyth_&_Stewart:Therm._Inf._Moduli_Prob.}.\footnote{Note here, that mild tuning 
($A<1\,$TeV) is needed for the quartic term due to the SUSY A-term to be 
ignored.} The non-renormalisable terms in \cref{Eq.: Full Potential0}
are the dominant terms in series over $\alpha$ and $n$. One would expect the 
lowest order to be dominant. Indeed, we find that parameter space exists only if
\mbox{$\alpha=n=1$}. Thus, we chose these values in this paper.\footnote{For a %
full study over all possible values of $\alpha$ and $n$ see \cite{My_Thesis}.}
With this choice, the potential in \cref{Eq.: Full Potential0} becomes
\begin{align}
V(\phi,\psi,T)=&V_{0}+ 
\left(g^2T^2-\frac12m_0^2+h^{2}\psi^{2}\right)\phi^{2}\nonumber\\
& +\lambda\frac{\phi^6}{M_{P}^{2}} + \frac{1}{2}m_{\psi}^{2}\psi^{2},
\label{Eq.: Full Potential}
\end{align}

We make the following definition
\begin{equation}
m^{2} \equiv m_{0}^{2} - 2h^{2}\psi^2\,.
\label{Eq.: Mass Definition}
\end{equation}
The variation of $m(\psi)$
is
\begin{equation}
\delta m=-\frac{2 h^{2}\psi}{m}\,\delta\psi\,.
\label{Eq.: Mass Variation}
\end{equation}

We only consider the case where the mass of $\phi$ is coupled to one field. Were
the mass coupled to several similar fields, the results would be just multiplied
by the number of fields. If the multiple fields are different, then there will 
be only a small number that dominate the contribution to the mass perturbation. 
Therefore we consider only one for simplicity.

Using \cref{Eq.: Mass Definition}, 
the potential becomes
\begin{align}
V(\phi,\psi,T)=&V_{0} + \left(g^{2}T^{2}-\frac{1}{2}m^{2}\right)\phi^{2}\nonumber\\
&+\lambda\frac{\phi^6}{M_{P}^{2}} + \frac{1}{2}m_{\psi}^{2}\psi^{2}\,.
\label{Eq.: Potential}
\end{align}
This potential is shown in \cref{Figure: Our Potential}.
\begin{figure}[h!]
\begin{center}
\includegraphics[scale=.8]{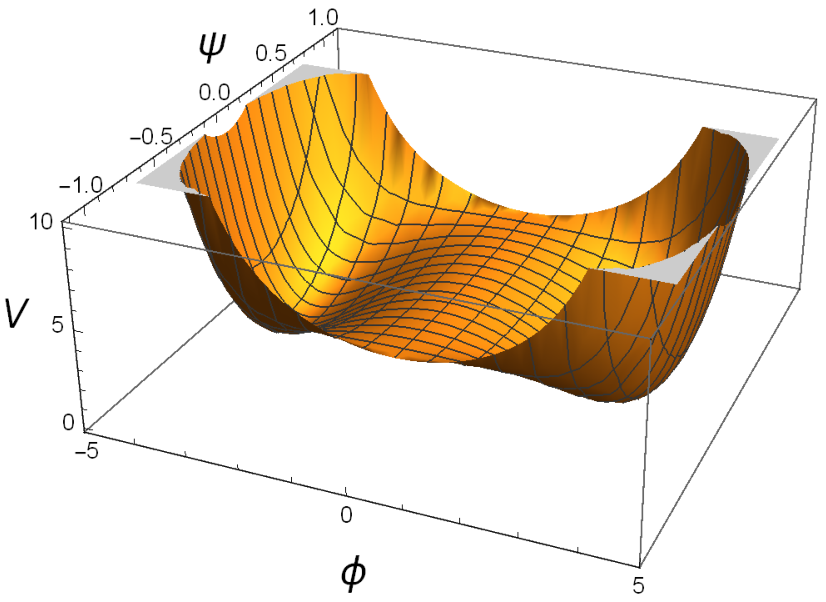}
\\
\textcolor{red}{Arbitrary Units}
\caption[Potential given by \cref{Eq.: Potential}]{The potential given by \cref{Eq.: Potential}.}
\label{Figure: Our Potential}
\end{center}
\end{figure}
It would appear from the potential that domain walls will be produced, due to 
the fact that in some parts of the Universe $\phi$ will roll down to 
$+\left\langle\phi\right\rangle$ while in others parts it will roll down to 
$-\left\langle\phi\right\rangle$. However, being a flaton field (i.e. a flat 
direction in SUSY) $\phi$ is a complex field, whose potential 
contains only one continuous vacuum expectation value (VEV).\footnote{A complex
$\phi$ may result in the copious appearance of cosmic strings after the end of 
thermal inflation. However, as their energy scale is very low (it is $V_0$),
they will not have any significant effect on the CMB observables. Moreover, 
depending on the overall background theory, such cosmic strings may well be 
unstable. Thus, we can safely ignore them.}

The zero temperature potential is
\begin{equation}
V(\phi,\psi,0)=
V_0-\frac12m^2\phi^2+\lambda\frac{\phi^6}{M_P^2}+\frac12m_\psi^2\psi^2\,.
\label{Eq.: 0 T Potential}
\end{equation}
Hence, the 
VEV is
\begin{equation}
\left\langle\phi\right\rangle\sim
\left(\frac{mM_{P}}{\sqrt{\lambda}}\right)^{1/2}.
\label{Eq.: Phi VEV}
\end{equation}
$V_{0}$ is obtained by requiring $V(\langle\phi\rangle)=0$
along the $\psi\!=\!0$ direction. We find
\begin{equation}
V_{0} \sim \frac{m_0^3M_{P}}{\sqrt\lambda}\,.
\label{Eq.: V_{0}}
\end{equation}
Now, we use the Friedmann equation
\begin{equation}
M_{P}^{2}H_{\textrm{TI}}^{2} \sim V_{0}\;,
\label{Eq.: V_{0} Energy Density Thermal Inflation}
\end{equation}
to obtain the Hubble parameter during thermal inflation as
\begin{equation}
H_{\textrm{TI}} \sim \left(\frac{m_{0}^{3}}{\sqrt{\lambda}\,M_{P}}\right)^{1/2}.
\label{Eq.: H_{TI}}
\end{equation}

Within this thermal inflation model there are two cases regarding the 
decay rate of the inflaton field $\Gamma_{\varphi}$, with $\varphi$ being the 
inflaton, i.e. the field driving primordial inflation prior to thermal 
inflation. One is the case when $\Gamma_{\varphi}\!\gtrsim\!H_{\textrm{TI}}$, 
i.e. that reheating from primordial inflation occurs before or around the time 
of the start of thermal inflation. Alternatively, there is the case when 
$\Gamma_{\varphi}\!\ll\!H_{\textrm{TI}}$, i.e. that reheating from primordial 
inflation occurs at some time after the end of thermal inflation. 

In the case of $\Gamma_{\varphi}\!\gtrsim\!H_{\textrm{TI}}$, 
thermal inflation begins at a temperature
\begin{equation}
T_{1} \sim V_{0}^{1/4}
\label{Eq.: T_{1} (Gamma_{varphi} gtrsim H_{TI})}
\end{equation}
$T_{1}$ corresponds to the temperature when the potential energy density becomes
comparable with the energy density of the thermal bath, for which the density 
is $\rho_{\gamma}\!\sim\!T^4$. 

In the case of $\Gamma_{\varphi}\!\ll\!H_{\textrm{TI}}$, thermal inflation begins at
a temperature\footnote{Before primordial reheating, the temperature is
\mbox{$T\sim \left(M_{P}^{2}H\Gamma_{\varphi}\right)^{1/4}$} \cite{KT}.}
\begin{equation}
T_{1} \sim \left(M_{P}^{2}\,H_{\textrm{TI}}\Gamma_{\varphi}\right)^{1/4}
\label{Eq.: T_{1} (Gamma_{varphi} << H_{TI})}
\end{equation}
Initially, for \mbox{$T\geq T_1$}, the thermal waterfall field is driven to zero
\mbox{$\phi\rightarrow 0$} as the thermally induced mass in 
\cref{Eq.: Potential} is dominant. This continues even if \mbox{$T<T_1$} as long
as the mass squared of $\phi$ remains positive. When the tachyonic mass term of
the thermal waterfall field becomes equal to the thermally-induced mass term
(cf. \cref{Eq.: Potential}) a phase transition sends the field towards its 
non-zero true VEV and thermal inflation ends 
\cite{Lyth_&_Stewart:Cos._TeV_Mass_Higgs_Field_break._GUT_Gauge._Sym.}. 

In both of the above cases, thermal inflation ends at a temperature
\begin{equation}
T_{2} = \frac{m}{\sqrt{2}\,g}\,.
\label{Eq.: T_{2}}
\end{equation}
In the following, we only consider the case where 
$\Gamma_{\varphi}\!\ll\!H_{\textrm{TI}}$, in that reheating from primordial 
inflation occurs at some time after the end of thermal inflation, as this 
scenario was found to yield more parameter space than the case where 
$\Gamma_{\varphi}\!\gtrsim\!H_{\textrm{TI}}$.

\section{\boldmath 
$\phi$ Decay Rate, Spectral Index and Tensor Fraction}
\label{Section: phi Decay Rate Spectral Index and Tensor Fraction}

\subsection{\boldmath $\phi$ Decay Rate}
\label{Subsection: phi Decay Rate}

The decay rate of $\phi$ is given by
\begin{equation}
\Gamma \sim \mathrm{max}\left\{g^{2}m \hspace{0.5em},\hspace{0.5em} \frac{m^{3}}{M_{P}^{2}}\right\}
\label{Eq.: Phi Decay Rate (Both Decay Channels)}
\end{equation}
The first expression is for decay into the thermal bath via direct interactions
and the second is for gravitational decay. We will only consider the case in 
which the direct decay is the dominant channel ($g$ is not taken to be very 
small). This is the case when \mbox{$m \ll gM_{P}$}. Therefore we have just
\mbox{$\Gamma \sim g^{2}m$}.

\subsection{Spectral Index 
and its running}
\label{Subsection: Spectral Index --- n_{s} and n_{s}^{'}}

Thermal Inflation has the effect of changing the number of e-folds before the 
end of primordial inflation at which cosmological scales exit the horizon. This 
affects the value of the spectral index $n_{s}$ of the curvature perturbation 
$\zeta$ (see for example \cite{charlotte,charlotte+}). 

We assume $\zeta$ is generated due to the perturbations of the spectator scalar 
field. Then, in the case of slow-roll inflation, the spectral index is given by 
\cite{Lyth_&_Liddle:Prim._Den._Pert.}
\begin{equation}
n_{s} \simeq 1 - 2\epsilon + 2\eta_{\psi}\;.
\label{Eq.: Spectral Index - Spectral Index Section}
\end{equation}
where $\epsilon$ and $\eta_{\psi}$ are slow-roll parameters, defined as
\begin{equation}
\epsilon\equiv\frac{M_{P}^2}{2}\left(\frac{V'(\varphi)}{V(\varphi)}\right)^{2}
\quad{\rm and}\quad 
\eta_{\psi}\equiv\frac{1}{3H^{2}}
\frac{\partial^{2}V}{\partial \psi^{2}}\,,
\label{Eq.:  epsilon and eta_{psi psi} Definition}
\end{equation}
where $V^{\prime}(\varphi)$ is the derivative of the inflaton potential with 
respect to the inflaton field $\varphi$. 
$\epsilon$ and $\eta_{\psi}$ are to be evaluated at the point where 
cosmological scales exit the horizon during primordial inflation. 



Regarding the various scalar fields involved in this model, the reason why 
$\epsilon$ depends only on $\varphi$ is because this slow-roll parameter 
captures the inflationary dynamics of primordial inflation, which is governed 
only by $\varphi$ in our model (we are assuming that both $\psi$ and $\phi$ 
have settled to a constant value (\cref{Subsubsection: 
The Field Value phi_{*},Subsubsection: The Field Value psi_{*}} respectively) 
by the time cosmological scales exit the horizon during primordial inflation). 
In a similar fashion, the reason why the slow-roll parameter $\eta_\psi$ depends
only on $\psi$ is because this parameter captures the dependence on the 
spectral index of the field(s) whose perturbations contribute to the observed 
primordial curvature perturbation $\zeta$. In our case, this is only the 
spectator field $\psi$.

The definition of the running of the spectral index is 
\cite{Lyth_&_Liddle:Prim._Den._Pert.}
\begin{equation}
n_{s}^{\prime} \equiv \frac{\mathrm{d}n_{s}}{\mathrm{d}\ln k}
\simeq -\frac{\mathrm{d}n_{s}}{\mathrm{d}N}\,,
\label{Eq.: Running of Spectral Index Definition}
\end{equation}
the second equation coming from $\mathrm{d}\ln k\!=\!\mathrm{d}\ln{(aH)}\!
\simeq\!H\mathrm{d}t\!\equiv\!-\mathrm{d}N$, where $k\!=\!aH$. From 
\cref{Eq.: Spectral Index - Spectral Index Section}, we have
\begin{equation}
n_{s}^{\prime}\simeq 2\frac{\mathrm{d}\epsilon}{\mathrm{d}N} 
- 2\frac{\mathrm{d}\eta_{\psi}}{\mathrm{d}N}
\simeq 2\epsilon\frac{\mathrm{d}\ln\epsilon}{\mathrm{d}N}
 - 2\frac{\mathrm{d}\eta_{\psi}}{\mathrm{d}N}\,.
\label{Eq.: Running of Spectral Index Working}
\end{equation}
Now, we have \cite{Lyth_&_Liddle:Prim._Den._Pert.}
\begin{equation}
\frac{\mathrm{d}\ln\epsilon}{\mathrm{d}N} \simeq -4\epsilon + 2\eta\,,
\label{Eq.: d ln(epsilon)/dN}
\end{equation}
where $\eta$ is a slow-roll parameter given by
\begin{equation}
\eta \equiv M_{P}^{2}\frac{V^{\prime\prime}(\varphi)}{V(\varphi)}\,.
\label{Eq.: eta Definition}
\end{equation}
Also,
\begin{equation}
\frac{\mathrm{d}\eta_{\psi}}{\mathrm{d}N} 
= -\frac{1
}{3H^{4}}\frac{\mathrm{d}(H^{2})}{\mathrm{d}N}
\frac{\partial^{2}V}{\partial \psi^{2}}
= -2\eta_{\psi}\frac{\mathrm{d}\ln H}{\mathrm{d}N}\,,
\label{Eq.: d eta_{psi psi}/dN Working}
\end{equation}
where we used that $V(\psi)$ does not depend on $N$, as we are assuming that 
both $\psi$ and $\phi$ have settled to a constant value (\cref{Subsubsection: 
The Field Value phi_{*},Subsubsection: The Field Value psi_{*}} respectively) 
by the time cosmological scales exit the horizon during primordial inflation. 

Since
\cite{Lyth_&_Liddle:Prim._Den._Pert.},
\begin{equation}
\frac{\mathrm{d}\ln H}{\mathrm{d}N} \simeq \epsilon\,,
\label{Eq.: d ln(H)/dN}
\end{equation}
we have
\begin{equation}
\frac{\mathrm{d}\eta_{\psi}}{\mathrm{d}N} \simeq -2\epsilon\eta_{\psi}\;.
\label{Eq.: d eta_{psi psi}/dN}
\end{equation}
Therefore, the final result for the running of the spectral index is
\begin{equation}
n_{s}^{\prime} \simeq -8\epsilon^{2} + 4\epsilon\eta + 4\epsilon\eta_{\psi}\;.
\label{Eq.: Running of Spectral Index - Spectral Index Section}
\end{equation}

From now on we assume that $H$ has the constant value $H_{\ast}$ by the time 
cosmological scales exit the horizon up until the end of primordial inflation. 
In order to obtain $\epsilon$ and $\eta$, we require the value of $N_{\ast}$, 
the number of e-folds before the end of primordial inflation at which 
cosmological scales exit the horizon. We consider the period between when the 
pivot scale, $k_0\!\equiv\!0.002\ \mbox{Mpc$^{-1}$}$, exits the horizon during 
primordial inflation and when it reenters the horizon long after the end of 
thermal inflation. We have
\begin{equation}
R_{\ast} = H_{\ast}^{-1} \quad{\rm and}\quad 
(k_0/a_{\rm piv})^{-1} = H_{\textrm{piv}}^{-1}\;,
\label{Eq.: R{*} & R_{pivot}}
\end{equation}
where $R_{\ast}$ is a length scale when the pivot scale exits the horizon during
primordial inflation and the subscript `piv' denotes the time when this scale 
re-enters the horizon, with $a$ being the scale factor of the Universe. 
Therefore
\begin{equation}
H_{\ast}^{-1} = \frac{a_{\ast}}{a_{\textrm{piv}}}H_{\textrm{piv}}^{-1}.
\label{Eq.: H_{*}^{-1}}
\end{equation}
Using the above, we now can calculate $N_*$.

Since $\Gamma_{\varphi}\!\ll\!H_{\textrm{TI}}$, we have
\begin{equation}
\hspace{-1cm}
e^{N_{\ast}}\!\!=\!\! \frac{H_{\ast}}{k}\!
\left(\frac{T_{\textrm{start,TI}}}{T_{\textrm{end,inf}}}\right)^{\!\!8/3}\!\!
\left(\frac{T_{\textrm{reh,TI}}}{T_{\textrm{end,TI}}}\right)^{\!\!8/3}\!\!
\frac{T_{\textrm{piv}}e^{-N_{\textrm{TI}}}}{T_{\textrm{reh,TI}}}
\hspace{-1cm}
\label{Eq.: e^{N_{*}} (Gamma_{varphi} << H_{TI})}
\end{equation}
where $N_{\textrm{TI}}$ is the number of e-folds of thermal inflation and the 
subscripts denote
the following: `end,inf' 
is at the end of primordial inflation, 
`start,TI' 
is at the start of thermal inflation, `end,TI' 
is at the end of thermal inflation and `reh,TI' 
is at thermal inflation reheating. 
For the period between the end of primordial/thermal inflation and 
primordial/thermal inflation reheating, \mbox{$a\propto{T}^{-8/3}$}.%
\footnote{
During this time, $T\!\sim\!\left(M_P^2H\Gamma_\varphi\right)^{1/4}$ \cite{KT}.
As \mbox{$H\propto t^{-1}$} we have $T\!\propto\!t^{-1/4}$. During the field 
oscillations, the Universe is matter dominated and so we have 
$a\!\propto\!t^{2/3}$. 
Putting this all together we find $T \propto t^{-1/4} \propto a^{-3/8}$.}
For all other times, $a\propto T^{-1}$. 

We need to calculate $T_{\textrm{piv}}$. We consider the period between when the pivot scale reenters the horizon and the present. Throughout this period the Universe is matter-dominated (ignoring dark energy). Therefore we have
\mbox{$\rho\propto a^{-3}\propto T^{3}$}.
Using the Friedmann equation, \mbox{$3M_{P}^{2}H^{2} \propto T^{3}$} we have
\begin{equation}
\frac{H_{\textrm{piv}}^{2}}{H_{0}^{2}} = \frac{T_{\textrm{piv}}^{3}}{T_{0}^{3}}
\;\Rightarrow\;
T_{\textrm{piv}} 
= 9.830\times10^{-13}\ \mbox{GeV}
\label{Eq.: T pivot Value}
\end{equation}
where `0' denotes the values at present.
%
%
Using this,
we obtain $N_{\ast}$ as
\begin{multline}
N_{\ast} \approx 
\ln{\left(\frac{\left(3.2\times10^{37}\ \mbox{GeV$^{-1}$}\right)H_{\ast}}%
{0.002}\right)} + \frac{2}{3}\ln{\left(\frac{\Gamma}{H_{\ast}}\right)}\\+ 
\frac{1}{4}\ln{\left(\frac{10\pi^{2}\left(9.8\times10^{-13}\ 
\mbox{GeV}\right)^{4}}{9M_{P}^{2}\Gamma^{2}}\right)} - N_{\textrm{TI}}\,,
\label{Eq.: N_{*} (Gamma_{varphi} << H_{TI})}
\end{multline}
where we have used $g_{\ast}\!\approx\!10^{2}$ as the number of spin states 
(effective relativistic degrees of freedom) of the particles in the thermal 
bath, at the time of both primordial inflation reheating and thermal inflation 
reheating.\footnote{Eq.~\eqref{Eq.: N_{*} (Gamma_{varphi} << H_{TI})}
is only valid as long as \mbox{$N_{\rm TI}>0$}. Otherwise $N_*$ is independent 
of $\Gamma$.}

\subsection{\boldmath Tensor Fraction $r$}
\label{Subsection: Tensor Fraction r}

The definition of the tensor fraction is
\mbox{$r \equiv \mathcal{P}_{h}/\mathcal{P}_{\zeta}\;$}
\cite{Lyth_&_Liddle:Prim._Den._Pert.}, 
where $\mathcal{P}_{h}$ and $\mathcal{P}_{\zeta}$ are the spectra of the 
primordial tensor and curvature perturbations respectively. The spectrum 
$\mathcal{P}_{h}$ is given by
\begin{equation}
\mathcal{P}_{h}(k) = \frac{8}{M_{P}^{2}}\left(\frac{H_{k}}{2\pi}\right)^{2}
\label{Eq.: P_{h}(k)}
\end{equation}
for a given wavenumber $k$. Using this, together with 
$\rho_{\ast}\!=\!3M_{P}^{2}H_{\ast}^{2}$, given that we are saying 
$H_{k}\!=\!H_{\ast}$ for our current case, as well as the observed value 
$\mathcal{P}_{\zeta}(k_{0})\!=\!2.142\times10^{-9}$, we obtain
\begin{equation}
r = \left(\frac{\rho_{\ast}^{1/4}}{3.25\times10^{16}\ \mbox{GeV}}\right)^{4}
\label{Eq.: r}
\end{equation}

\section{End-of-Inflation Mechanism}
\label{Section: ``End of Inflation'' Mechanism}

In this section we investigate the ``end of inflation'' mechanism. We aim to 
obtain a number of constraints on the model parameters and the initial 
conditions for the fields. Considering these constraints, we intend to determine
the available parameter space. In this parameter space we will calculate 
distinct observational signatures that may test this scenario in the near 
future.\footnote{We also investigated a modulated decay rate scenario, but found that there was no parameter space available. For our detailed work on this, see \cite{My_Thesis}.}

\subsection{\boldmath Generating $\zeta$}
\label{Subsection: Generating zeta}

As $\phi$ is coupled to $\psi$, the ``end of inflation'' mechanism will 
generate a contribution to the primordial curvature perturbation $\zeta$ 
\cite{Lyth:Generating_Curvature_Perturbation_E._of_I.}. We use the $\delta N$ 
formalism to calculate this contribution
as
\begin{equation}
\zeta\! =\! \delta N_{\textrm{TI}} = \!
\frac{\mathrm{d}N_{\textrm{TI}}}{\mathrm{d}m}\delta m + 
\frac{1}{2!}\frac{\mathrm{d^{2}}N_{\textrm{TI}}}{\mathrm{d}m^{2}}\delta m^{2} + 
\frac{1}{3!}\frac{\mathrm{d^{3}}N_{\textrm{TI}}}{\mathrm{d}m^{3}}\delta m^{3} + 
\cdots\,.
\label{Eq.: zeta Definition - Our Work}
\end{equation}
The number of e-folds between the start and end of thermal inflation is given by
\begin{equation}
N_{\textrm{TI}} = 
\ln{\left(\frac{a_2}{a_1}\right)} = \ln{\left(\frac{T_1}{T_2}\right)}\,,
\label{Eq.: N}
\end{equation}
where $a_{1}\!=\!a_{\textrm{start,TI}}$ and $a_{2}\!=\!a_{\textrm{end,TI}}$.

Substituting $T_{1}$ and $T_{2}$, 
\cref{Eq.: T_{1} (Gamma_{varphi} << H_{TI}),Eq.: T_{2}} respectively, into 
\cref{Eq.: N} gives
\begin{align}
N_{\textrm{TI}} &\simeq 
\ln{\left[
\frac{\sqrt{2}\,g\left(M_{P}^{2}\,H_{\textrm{TI}}\Gamma_{\varphi}\right)^{1/4}}{m}
\right]}\nonumber\\
&\simeq
\frac18\ln\left(\frac{g^8}{\sqrt\lambda}\frac{M_P^3\Gamma_\varphi^2}{m^5}\right),
\label{Eq.: N (Gamma_{varphi} << H_{TI})}
\end{align}
where we used \cref{Eq.: H_{TI}}

Therefore the $\delta N$ formalism to third order gives
\begin{equation}
\zeta = \delta N_{\textrm{TI}} = 
-\frac58\frac{\delta m}{m} + \frac{5}{16}\frac{\delta m^{2}}{m^{2}} 
- \frac{5}{24}\frac{\delta m^{3}}{m^{3}}\,.
\label{Eq.: Zeta (m) (Gamma_{varphi} << H_{TI})}
\end{equation}
By substituting our mass definition and its differential, 
\cref{Eq.: Mass Definition,Eq.: Mass Variation}, into 
\cref{Eq.: Zeta (m) (Gamma_{varphi} << H_{TI})} we obtain the power spectrum of
the primordial curvature perturbation,\footnote{%
It must be noted that although there will be perturbations in $\psi$ that are 
generated during thermal inflation that will become classical due to inflation, 
the scales to which these correspond are much smaller than cosmological scales, 
as thermal inflation lasts for only about 10-15 e-folds. 
Therefore we do not consider them here.} which to first order is
\begin{equation}
\sqrt{\mathcal{P}_{\zeta}} = \frac{5}{8\pi}
\frac{h^{2}H_{\ast}\psi}{m^{2}}\,.
\label{Eq.: Power Spectrum (Gamma_{varphi} << H_{TI})}
\end{equation}

A required condition for the perturbative expansion in 
\cref{Eq.: Zeta (m) (Gamma_{varphi} << H_{TI})} to be suitable is that each
term is much smaller than the preceding one. This requirement gives
\begin{equation}
\frac{h^{2}H_{\ast}\psi}{m^{2}} \ll 1\,,
\label{Eq.: Suitable Perturbative Expansion}
\end{equation}
which is readily satisfied as $\sqrt{\mathcal{P}_{\zeta}}\ll 1$.

\subsection{Constraining the 
Parameters}
\label{Subsection: Constraining the Free Parameters - End of Inflation Section}

In this section we produce a number of constraints for the model parameters 
and we describe the rationale behind them.

\subsubsection{Primordial Inflation Energy Scale}
\label{Subsubsection: Primordial Inflation Energy Scale}

We want the energy scale of primordial inflation to be
\mbox{$V^{1/4} \lesssim 10^{14}\,$GeV} so that the inflaton contribution 
to the curvature perturbation is negligible. Therefore, from the Friedmann 
equation 
we require
\begin{equation}
H_{\ast} \lesssim 10^{10}\ \mbox{GeV}
\label{Eq.: H_{*} Constraint a}
\end{equation}

\subsubsection{Thermal Inflation Dynamics}
\label{Subsubsection: Thermal Inflation Dynamics}

We will consider only the case in which the inflationary trajectory is 
1-dimensional, in that only the $\phi$ field is involved in determining the 
trajectory of thermal inflation in field space. We do this only to work with 
the simplest scenario for the trajectory. It is not a requirement on the model 
itself. In order that the $\psi$ field does not affect the inflationary 
trajectory during thermal inflation, we require from our $m$ mass definition, 
\cref{Eq.: Mass Definition}, that
\begin{equation}
m_{0}\gtrsim h\psi\,.
\label{Eq.: m_{0} >> Coupling Term}
\end{equation}
Therefore we have
$m \simeq m_{0}\;$.

From our potential, at the onset of thermal inflation,
\cref{Eq.: Full Potential}, \cref{Eq.: m_{0} >> Coupling Term} gives
\begin{equation}
m_{0}^{2} < 2g^{2}T_{1}^{2}.
\label{Eq.: m_{0}^{2}<2g^{2}T_{1}^{2}}
\end{equation}
For $\Gamma_{\varphi}\!\ll\!H_{\textrm{TI}}$, substituting $T_{1}$ from 
\cref{Eq.: T_{1} (Gamma_{varphi} << H_{TI})} into 
\cref{Eq.: m_{0}^{2}<2g^{2}T_{1}^{2}} gives
\begin{equation}
m_0<
\left[\left(g^4\Gamma_\varphi\right)^2\frac{M_P^3}{\sqrt{\lambda}}\right]^{1/5}.
\label{Eq.: m_{0} Constraint c (Gamma_{varphi} << H_{TI})}
\end{equation}

\subsubsection{\boldmath Lack of Observation of $\phi$ Particles}
\label{Subsubsection: Lack of Observation of phi Particles}

Given that we have not observed any $\phi$ particles, the 
constraint on the present value of the effective mass of $\phi$ is
\mbox{$m_{\phi,\textrm{now}} \gtrsim 1\,$TeV}.
From our potential, \cref{Eq.: Full Potential}, we have
\mbox{$m_{\phi,\textrm{now}}^{2} \sim -m_{0}^{2} + 
30\lambda\left\langle\phi\right\rangle^4/M_{P}^{2}$}.
Substituting the VEV of $\phi$, \cref{Eq.: Phi VEV}, into here gives
\mbox{$m_{\phi,\textrm{now}} \sim m_{0}$}
for all reasonable values of $n$. Therefore, we require
\begin{equation}
m_{0} \gtrsim 1\ \mbox{TeV}\,.
\label{Eq.: m_{0} gtrsim 1 TeV}
\end{equation}

\subsubsection{\boldmath Light auxiliary field $\psi$}
\label{Subsubsection: Light psi - End of Inflation Subsubsection}

In order that $\psi$ acquires classical perturbations during primordial 
inflation, we require $\psi$ to be light during this time, i.e. 
\mbox{$|m_{\psi,\textrm{eff}}| \ll H_{\ast}$}, where we are using notation such that 
\mbox{$|m_{\psi,\textrm{eff}}|\equiv\sqrt{\left|m_{\psi,\textrm{eff}}^{2}\right|}$}. 
We have
\begin{equation}
m_{\psi,\textrm{eff}}^{2} = m_{\psi}^{2} + 
2h^{2}\phi^{2}\,.
\label{Eq.: Psi Effective Mass Squared - 1st appearance}
\end{equation}
Therefore we need
\begin{equation}
m_{\psi}<H_{\ast}
\quad{\rm and}\quad
h\phi_{\ast}
<H_{\ast}\;,
\label{Eq.: 2nd Term Psi Effective Mass << H_{*}}
\end{equation}
where $\phi_{\ast}$ and $\psi_{\ast}$ are the values of $\phi$ and $\psi$ during 
primordial inflation respectively.

We require that $\psi$ remains at $\psi_{\ast}$, the value during primordial 
inflation, all the way up to the end of thermal inflation. The reason for this 
is that if $\psi$ started to move, then its perturbation would decrease. This 
is because $\psi$ unfreezes when the Hubble parameter becomes less than $\psi$'s
mass, i.e. $H\!<\!m_{\psi}$. In this case, the perturbation of $\psi$ also 
unfreezes, because it has the same mass as $\psi$. The density of the 
oscillating $\psi$ field decreases as matter, so 
\mbox{$m_{\psi}^{2}\psi^{2}\!\propto\!a^{-3}\Rightarrow\psi\!\propto\!a^{-3/2}$}. 
The same is true for the perturbation, i.e. $\delta\psi\!\propto\!a^{-3/2}$. 
So the whole effect of perturbing the end of thermal inflation is diminished. 
Requiring that $\psi$ is light at all times up until the end of thermal 
inflation is sufficient to ensure that the field and its perturbation remain at 
$\psi_{\ast}$ and $\delta\psi_{\ast}$ respectively. Therefore we require
\begin{equation}
m_{\psi}<H_{\textrm{TI}}
\label{Eq.: Psi Mass << H_{TI}}
\end{equation}
which is of course stronger than just \mbox{$m_{\psi}\!\ll\!H_{\ast}$} 
in \cref{Eq.: 2nd Term Psi Effective Mass << H_{*}}.

Similarly to $\phi$, given that we have not observed any $\psi$ particles, the 
most liberal constraint on the present value of the effective mass of $\psi$ is
\begin{equation}
m_{\psi,\textrm{now}} \gtrsim 1\ \mbox{TeV}
\label{Eq.: m_{psi,now} gtrsim 1 TeV}
\end{equation}

\subsubsection{\boldmath The Field Value $\psi_{\ast}$}
\label{Subsubsection: The Field Value psi_{*}}

Substituting the observed spectrum value 
$\mathcal{P}_{\zeta}(k_{0})\!=\!2.142\times10^{-9}$ into 
\cref{Eq.: Power Spectrum (Gamma_{varphi} << H_{TI})} 
gives the constraint
\begin{equation}
\psi_{\ast} \sim 10^{-4}\,
\frac{m_{0}^{2}}{h^{2}H_{\ast}}\,.
\label{Eq.: Psi_{*}}
\end{equation}
%
Substituting \cref{Eq.: Psi_{*}} into \cref{Eq.: m_{0} >> Coupling Term}, 
regarding the dynamics of thermal inflation, gives
\begin{equation}
h \gtrsim 10^{-4}\,\frac{m_{0}}{H_{\ast}}\,.
\label{Eq.: h Constraint h}
\end{equation}
Rearranging this for $m_{0}$ gives the constraint
\begin{equation}
m_{0} \lesssim 10^{4}\,hH_{\ast}\;.
\label{Eq.: m_{0} Constraint h}
\end{equation}

We require the field value of $\psi$ to be much larger than its perturbation, 
i.e. $\psi_{\ast}\gg\delta\psi_{\ast}$, so that the perturbative approach is valid. Therefore, with $\delta\psi_{\ast} \sim H_{\ast}$, we obtain
\begin{equation}
\psi_{\ast} \gg H_{\ast}
\quad{\rm and}\quad
\frac{\delta\psi_{\ast}}{\psi_{\ast}} \ll 1\,.
\end{equation}
Combining the frozen value $\psi_{\ast}$, \cref{Eq.: Psi_{*}}, with the above
gives
\begin{equation}
m_{0}>10^2\,hH_{\ast}\;.
\label{Eq.: m_{0} Constraint e}
\end{equation}
Thus, we find the following range
\begin{equation}
10^2<\frac{m_0}{hH_*}<10^4.
\label{mrange}
\end{equation}

\subsubsection{\boldmath Thermal Fluctuation of $\phi$}
\label{Subsubsection: Thermal Fluctuation of phi}

The effective mass of $\phi$ at the end of primordial inflation is
\begin{equation}
m_{\phi,\textrm{end,inf}}^{2} \sim g^{2}T_{\textrm{end,inf}}^{2} - m_{0}^{2}
\sim g^2\,T_{\textrm{end,inf}}^2\;,
\label{Eq.: phi Effective Mass End Prim. Inf.}
\end{equation}
since $gT_{\textrm{end,inf}}\!\gg\!m_{0}$ \cite{My_Thesis}. 

As we are dealing with the thermal fluctuation of $\phi$ about $\phi\!=\!0$, 
we have $\left\langle\delta\phi\right\rangle_{T}\!=
\!\left\langle\phi\right\rangle_{T}$. The thermal fluctuation of $\phi$ is
\begin{equation}
\sqrt{\left\langle\phi^{2}\right\rangle_{T}} \sim T
\label{Eq.: phi Thermal Fluctuation}
\end{equation}
and we require \cite{My_Thesis}
\begin{equation}
g<1\,,
\label{Eq.: g << 1 (Therm. Fluc. phi Subsubsection)}
\end{equation}
because $g$ is a perturbative coupling.

In order to keep $m_{\psi,\textrm{eff}}$ light, we require 
(cf.\cref{Subsubsection: Light psi - End of Inflation Subsubsection}),
\begin{equation}
hT_{1}<H_{\textrm{TI}}\;.
\label{Eq.: Thermal Fluctuation Constraint a}
\end{equation}
During the time between the end of primordial inflation and primordial 
inflation reheating, $T\!\propto\!a^{-3/8}$ and $H\!\propto\!a^{-3/2}$. 
Therefore, 
if \cref{Eq.: Thermal Fluctuation Constraint a} is satisfied, then equivalent 
constraints for higher $T$ and $H$ are guaranteed to be satisfied as well. 

Considering $\Gamma_{\varphi}\!\ll\!H_{\textrm{TI}}$, by substituting 
\cref{Eq.: H_{TI},Eq.: Psi_{*},Eq.: T_{1} (Gamma_{varphi} << H_{TI})} 
into \cref{Eq.: Thermal Fluctuation Constraint a} we obtain the constraint
\begin{equation}
h<
\left(\lambda^{-3/2}\frac{m_0^9}{M_P^7\Gamma_\varphi^2}\right)^{1/8}
\label{Eq.: h Constraint e (Gamma_{varphi} << H_{TI})}
\end{equation}
Rearranging this for $m_{0}$ gives 
\begin{equation}
m_{0} > \left(\lambda^{3/2}h^8M_P^7\Gamma_\varphi^2\right)^{1/9}.
\label{Eq.: m_{0} Constraint i (Gamma_{varphi} << H_{TI})}
\end{equation}

\subsubsection{\boldmath Thermalization of $\phi$}
\label{Subsubsection: Thermalization of phi}

In order that $\phi$ interacts with the thermal bath and therefore that we 
actually have the $g^{2}T^{2}\phi^{2}$ term in our potential, 
\cref{Eq.: Full Potential}, we require \mbox{$\Gamma_{\textrm{therm}} > H$},
where $\Gamma_{\textrm{therm}}$ is the thermalization rate of $\phi$, which is 
given by 
\begin{equation}
\Gamma_{\textrm{therm}} = n\left\langle\sigma v\right\rangle
\sim \sigma\,T^{3}\,,
\end{equation}
where $n\!\sim\!T^{3}$ is the number density of particles in the thermal bath, 
$\sigma$ is the scattering cross-section for the interaction of $\phi$ and the 
particles in the thermal bath, $v$ is the relative velocity between a $\phi$ 
particle and a thermal bath particle (which in our case is $\approx\!c\!=\!1$) 
and $\left\langle\,\right\rangle$ denotes a thermal average. The scattering 
cross-section $\sigma$ is given by
\begin{equation}
\sigma \sim \frac{g^{4}}{E_{\textrm{cm}}^{2}}\,,
\label{Eq.: Cross-section -- Thermalization}
\end{equation}
where $E_{\textrm{cm}}$ is the centre-of-mass energy, which is
\mbox{$E_{\textrm{cm}} \sim T$}. Substituting this into 
\cref{Eq.: Cross-section -- Thermalization} gives
\begin{equation}
\sigma \sim \frac{g^{4}}{T^{2}}\,.
\label{Eq.: Cross-section Final Result -- Thermalization}
\end{equation}
This scattering cross-section is the total cross-section for all types of 
scattering (e.g. elastic) that can take place between $\phi$ and the particles 
in the thermal bath.\footnote{For a complete Field Theory derivation of the 
elastic scattering cross-section between $\phi$ and the thermal bath, see 
\cite{My_Thesis}.} The thermalization rate now becomes
\begin{equation}
\Gamma_{\textrm{therm}} \sim g^{4}T\,.
\label{Eq.: Thermalization Rate}
\end{equation}

As before, during the time between the end of primordial inflation and 
primordial inflation reheating, $T\!\propto\!a^{-3/8}$ and $H\!\propto\!a^{-3/2}$.
Therefore, if the constraint $\Gamma_{\textrm{therm}}\!>\!H$ is satisfied at the 
time of the end of primordial inflation, then it is satisfied all the way up to 
the start of thermal inflation. Thus, we have the constraint
\begin{equation}
\Gamma_{\textrm{therm}} \gtrsim H_{\ast}\,.
\label{Eq.: Gamma_{therm} > H_{*}}
\end{equation}

Taking \cref{Eq.: Thermalization Rate} with 
$T\!\sim\!\left(M_{P}^{2}\,H_{\ast}\,\Gamma_{\varphi}\right)^{1/4}$ gives
\begin{equation}
\Gamma_{\varphi}\gtrsim\frac{H_{\ast}^{3}}{g^{16}M_{P}^{2}}\,.
\label{Eq.: Gamma_{varphi} Constraint a}
\end{equation}
We also require $\Gamma_{\textrm{therm}}\!>\!H$ to be satisfied throughout the whole of thermal inflation. Therefore, we have the constraint
\begin{equation}
g^{4}\,T_{2} > H_{\textrm{TI}}\;.
\label{Eq.: g^{4}T_{2} > H_{TI}}
\end{equation}
Substituting $H_{\textrm{TI}}$ and $T_{2}$, \cref{Eq.: T_{2},Eq.: H_{TI}} into the 
above gives
\begin{equation}
m_{0} < 
g^6\sqrt{\lambda}\,M_{P}\;.
\label{Eq.: m_{0} Constraint u}
\end{equation}

\subsubsection{\boldmath The Field Value $\phi_{\ast}$}
\label{Subsubsection: The Field Value phi_{*}}

We consider two possible cases for the value of the thermal waterfall field 
$\phi$ during primordial inflation, with $m_{\phi,\textrm{inf}}$ being the 
effective mass of $\phi$ during primordial inflation:
\begin{enumerate}[label=\Alph*)]
\item $\phi$ heavy, i.e. $|m_{\phi,\textrm{inf}}| \gg H_{\ast}$, 
in which $\phi$ rolls down to its VEV.
\item $\phi$ light, i.e. $|m_{\phi,\textrm{inf}}| \ll H_{\ast}$, 
in which $\phi$ is at the Bunch-Davies value (to be explained below).
\end{enumerate}
\underline{\textbf{Case A}}\vspace{0.2em}
\\
Substituting $\left\langle\phi\right\rangle$, \cref{Eq.: Phi VEV}, into 
\cref{Eq.: 2nd Term Psi Effective Mass << H_{*}} gives
\begin{equation}
h< \lambda^{1/4}
\frac{H_{\ast}}{\sqrt{m_{0}M_{P}}}\,.
\label{Eq.: h Constraint f}
\end{equation}
Rearranging this for $m_{0}$ gives
\begin{equation}
m_{0}< 
\frac{\sqrt\lambda}{h^2}\,\frac{H_{\ast}^2}{M_{P}}\,.
\label{Eq.: m_{0} Constraint m}
\end{equation}
\\
\\
\\
\underline{\textbf{Case B}}\vspace{0.2em}
\\
We consider $\phi$ to be at the Bunch-Davies value
\begin{equation}
\phi_{\textrm{BD}} \sim \left(\frac{M_{P}H_{\ast}^{2}}{\sqrt{\lambda}}\right)^{1/3},
\label{Eq.: Phi_* B.-D. Value}
\end{equation}
corresponding to the Bunch-Davies vacuum 
\cite{Bunch_&_Davies:Quan._Field_Theory_de_Sit._Space:Renorm._Point_Splitting},
which is the unique quantum state that corresponds to the vacuum, i.e. no 
particle quanta, in the infinite past in conformal time in a de Sitter 
spacetime. $\phi_{\textrm{BD}}$ is of this form as 
$\lambda\phi^6/M_{P}^{2}\sim H_{\ast}^{4}$, this being because the 
probability of this Bunch-Davies state is proportional to the factor 
$e^{-V/H^{4}}$ \cite{staroyoko}. 

Substituting $\phi_{\textrm{BD}}$, \cref{Eq.: Phi_* B.-D. Value}, into 
\cref{Eq.: 2nd Term Psi Effective Mass << H_{*}} gives
\begin{equation}
h<\lambda^{1/6}\left(\frac{H_{\ast}}{M_{P}}\right)^{1/3}.
\label{Eq.: h Constraint g}
\end{equation}

\subsubsection{
\boldmath Energy Density of $\phi$}
\label{Subsubsection: Energy Density of the Thermal Waterfall Field}

We require the energy density of $\phi$ to be subdominant at all times, in order
that it does not cause any inflation by itself. During the period between the 
end of primordial inflation and the start of thermal inflation, the energy 
density of $\phi$ is
\begin{equation}
\rho_{\phi}\sim g^{2}T^{2}\phi^{2}\sim g^{2}T^{4},
\label{Eq.: phi Energy Density between End Primordial Inflation and 
Start Thermal Inflation}
\end{equation}
the second equation coming from the thermal fluctuation $\phi\sim\!T$. 
Therefore, considering the Friedmann equation, we require
\begin{equation}
g\,T_{1}^{2}<M_{P}H_{\textrm{TI}}\;.
\label{Eq.: gT_{1}^{2} << M_{P}H_{TI}}
\end{equation}
During the time between the end of primordial inflation and primordial inflation
reheating, $T\!\propto\!a^{-3/8}$ and $H\!\propto\!a^{-3/2}$. 
Therefore, if 
\cref{Eq.: gT_{1}^{2} << M_{P}H_{TI}} is satisfied, then equivalent constraints 
for higher $T$ and $H$ are guaranteed to be satisfied as well. 

Using that $\Gamma_{\varphi}\!\ll\!H_{\textrm{TI}}$, by substituting $H_{\textrm{TI}}$ 
and $T_{1}$, \cref{Eq.: T_{1} (Gamma_{varphi} << H_{TI}),Eq.: H_{TI}} 
into \cref{Eq.: gT_{1}^{2} << M_{P}H_{TI}} we obtain
\begin{equation}
m_0>\left[\left(g^2\Gamma_\varphi\right)^2\sqrt{\lambda}\,M_P\right]^{1/3}.
\label{Eq.: m_{0} Constraint w}
\end{equation}
\\
\underline{\textbf{\boldmath $\phi_{\ast}$ Case A}}\vspace{0.2em}
\\
The energy density of $\phi$ during primordial inflation is
\begin{align}
\rho_{\phi,\textrm{inf}} &= 
\left(-\frac{1}{2}m_{0}^{2} + h^{2}\psi_{\ast}^{2}\right)
\left\langle\phi\right\rangle^{2} + 
\lambda\frac{\left\langle\phi\right\rangle^6}{M_{P}^{2}}
\nonumber\\
&\sim -\frac{1}{2}m_{0}^{2}\left\langle\phi\right\rangle^{2} + 
\lambda\frac{\left\langle\phi\right\rangle^6}{M_{P}^{2}}\,,
\label{Eq.: Phi Ener. Den. Prim. Inf. - Phi_{*} Case A}
\end{align}
with the second equation coming from \cref{Eq.: m_{0} >> Coupling Term} 
regarding the dynamics of thermal inflation. Therefore, with the energy density 
of the Universe being $\sim\!M_{P}^{2}H_{\ast}^{2}$, we require
\begin{equation}
m_{0}\left\langle\phi\right\rangle < M_{P}H_{\ast}
\quad{\rm and}\quad
\sqrt{\lambda}\left\langle\phi\right\rangle^3 < M_{P}^2H_{\ast}\;.
\end{equation}
Substituting $\left\langle\phi\right\rangle$, \cref{Eq.: Phi VEV}, into 
the above gives the constraint 
\begin{equation}
m_{0} < \left(\sqrt{\lambda}\,M_{P}H_{\ast}^2\right)^{1/3}.
\label{Eq.: m_{0} Constraint o}
\end{equation}
\\
\underline{\textbf{\boldmath $\phi_{\ast}$ Case B}}\vspace{0.2em}
\\
The energy density of $\phi$ during primordial inflation is
\begin{align}
\rho_{\phi,\textrm{inf}} &= 
\left(-\frac{1}{2}m_{0}^{2} + h^{2}\psi_{\ast}^{2}\right)\phi_{\textrm{BD}}^{2} + 
\lambda\frac{\phi_{\textrm{BD}}^6}{M_{P}^{2}}\nonumber\\
&\sim 
-\frac12m_0^2\phi_{\textrm{BD}}^{2} + \lambda\frac{\phi_{\textrm{BD}}^6}{M_{P}^{2}},
\label{Eq.: Phi Ener. Den. Prim. Inf. - Phi_{*} Case B}
\end{align}
with the second equation coming from \cref{Eq.: m_{0} >> Coupling Term} 
regarding the dynamics of thermal inflation. Therefore, with the energy density
of the Universe being $\sim\!M_{P}^{2}H_{\ast}^{2}$, we require
\begin{equation}
m_{0}\,\phi_{\textrm{BD}} < M_{P}H_{\ast}
\quad{\rm and}\quad
\sqrt{\lambda}\,\phi_{\textrm{BD}}^3 < M_{P}^2H_{\ast}
\end{equation}
Substituting $\phi_{\textrm{BD}}$, \cref{Eq.: Phi_* B.-D. Value}, into the above
gives
\begin{equation}
m_{0} < \left(\sqrt{\lambda}\,M_{P}^{2}H_{\ast}\right)^{1/3}.
\label{Eq.: m_{0} Constraint p}
\end{equation}

\subsubsection{Transition from Thermal Inflation to 
Thermal Waterfall Field Oscillation}
\label{Subsubsection: Time of Transition from Thermal Inflation to 
Thermal Waterfall Field Oscillation}

In order for the equations of the $\delta N$ formalism that are derived within 
the context of the ``end of inflation'' mechanism to be valid, we require the 
transition from thermal inflation to thermal waterfall field oscillation to be 
sufficiently fast \cite{Lyth:Hybrid_Waterfall_and_Curvature_Perturbation}. More 
specifically, we require
\begin{equation}
\Delta t < \delta t_{1\rightarrow2}\;,
\label{Eq.: Delta t << delta t_{1->2}}
\end{equation}
where $\Delta t\!\equiv\!t_{2}-t_{1}$ is the time taken for the transition to 
occur and $\delta t_{1\rightarrow2}$ is the proper time between a uniform energy 
density spacetime slice just before the transition at $t_{1}$ and one just after
the transition at $t_{2}$ when $\phi$ starts to oscillate around its VEV. 
Qualitatively, we require the thickness of the transition slice to be much 
smaller than its warping. 

The primordial curvature perturbation that is generated by the 
``end of inflation'' mechanism is 
\begin{equation}
\zeta = H_{\textrm{TI}}\, \delta t_{1\rightarrow2}\;.
\label{Eq.: zeta = H delta t_{1->2}}
\end{equation}
Therefore, from \cref{Eq.: Delta t << delta t_{1->2}} we require
\begin{equation}
\zeta > H_{\textrm{TI}} \Delta t\,.
\label{Eq.: zeta >> H Delta t}
\end{equation}

To calculate $\phi_{1}$ and $\phi_{2}$, the value of $\phi$ at times $t_{1}$ and 
$t_{2}$ respectively, we use the fact that the process is so rapid that it takes
place in less than a Hubble time, so that the Universe expansion can be ignored.
Then the equation of motion is
\begin{equation}
\ddot{\phi} + \frac{\partial V}{\partial \phi} \simeq 0\,.
\label{Eq.: Fast-roll EoM}
\end{equation}
At the end of thermal inflation, $\phi$ is not centred on the origin, but has 
started to roll down the potential slightly. At this time, $g^{2}T^{2}$ is much 
smaller than $m_{0}^{2}$. Therefore we have
\begin{equation}
\frac{\partial V}{\partial \phi} \simeq -m_{0}^{2}\phi\,.
\label{Eq.: V' at end of inflation}
\end{equation}
So we have the equation of motion
\mbox{$\ddot{\phi} \simeq m_{0}^{2}\phi$}
whose solution is
\begin{equation}
\phi\propto e^{m_{0}t},
\label{Eq.: Fast-roll Phi Solution}
\end{equation}
where 
we are considering only the growing mode. 
Therefore we have
\begin{equation}
\ln{\left(\frac{\phi_{2}}{\phi_{1}}\right)} \sim m_{0}\left(t_{2}-t_{1}\right)
\sim m_{0} \Delta t\,.
\label{Eq.: ln(Phi_{2}/Phi_{1})}
\end{equation}
We know that \mbox{$\phi_{1} \sim T \sim m_{0}$} and
\mbox{$\phi_{2} \sim \left\langle\phi\right\rangle$}.
Therefore we have
\begin{equation}
\ln{\left[\left(\frac{1}{\sqrt\lambda}
\frac{M_{P}}{m_{0}}\right)^{1/2}\right]} \sim m_{0} \Delta t\,.
\label{Eq.: ln() sim m_{0} Delta T Final}
\end{equation}
For all values of $\lambda$ and $m_{0}$, we have $\Delta t\!\geq\!m_{0}^{-1}$. 
Therefore, from \cref{Eq.: zeta >> H Delta t} we have
\begin{equation}
\zeta > \frac{H_{\textrm{TI}}}{m_{0}}\,.
\label{Eq.: zeta >> H/m_{0}}
\end{equation}
Thus, given that $\zeta\sim10^{-5}$, we require
\begin{equation}
H_{\textrm{TI}} < 10^{-5}m_{0}\,.
\label{Eq.: H_{TI} Constraint e}
\end{equation}

We obtain an additional constraint by substituting 
\cref{Eq.: H_{TI} Constraint e} into the requirement of 
$m_{\psi}\!\ll\!H_{\textrm{TI}}$, \cref{Eq.: Psi Mass << H_{TI}}. This gives
\begin{equation}
m_{\psi} < 10^{-5}m_{0}\,.
\label{Eq.: Psi Mass Constraint a}
\end{equation}
A further constraint is obtained by substituting $H_{TI}$, \cref{Eq.: H_{TI}}, into \cref{Eq.: H_{TI} Constraint e}. We obtain
\begin{equation}
m_{0} < 10^{-10}\sqrt{\lambda}\,M_{P}\;.
\label{Eq.: m_{0} Constraint f}
\end{equation}

\subsubsection{\boldmath Energy Density of the Oscillating $\psi$}
\label{Subsubsection: Energy Density of the Oscillating Spectator Field}

As $\psi$ has acquired perturbations from primordial inflation, we require it 
not to dominate the energy density of the Universe after the end of thermal 
inflation when it is oscillating, at which time the effective mass of $\psi$ is
increased significantly due to the coupling of $\psi$ to $\phi$. This is so as 
not to allow $\psi$ to act as a curvaton, i.e. not to allow $\psi$'s 
perturbations to generate a significant contribution to the primordial curvature
perturbation when $\psi$ decays. The reason for this is just so that 
we do not have a curvaton inflation scenario, where the perturbations generated
via the modulated mass give a negligible contribution to $\zeta$.

The energy density of the oscillating $\psi$ field after the end of thermal 
inflation is
\begin{equation}
\rho_{\psi,\textrm{osc}} = h^{2}\overline{\psi^{2}}\,\overline{\phi^{2}} + 
\frac{1}{2}m_{\psi}^{2}\overline{\psi^{2}}
\sim h^{2}\psi_{\ast}^{2}\left\langle\phi\right\rangle^{2} + 
\frac{1}{2}m_{\psi}^{2}\psi_{\ast}^{2}\,.
\label{Eq.: Psi Energy Density Oscillating}
\end{equation}
For simplicity, we assume that $\psi$ decays around the same time as $\phi$, 
i.e. that $H$ does not change much between the time when $\phi$ decays and the 
time when $\psi$ decays. Therefore, the energy density of the Universe at the 
time when $\psi$ decays is $\sim\!M_{P}^{2}\Gamma^{2}$. We therefore require
\begin{equation}
\rho_{\psi,\textrm{osc}} \sim
h^{2}\psi_{\ast}^{2}\left\langle\phi\right\rangle^{2} + 
\frac{1}{2}m_{\psi}^{2}\psi_{\ast}^{2} < M_{P}^{2}\Gamma^{2},
\label{Eq.: Psi Energy Density Oscillating Constraint a}
\end{equation}
which means
\begin{equation}
m_{\psi} < \frac{M_{P}\Gamma}{\psi_{\ast}}
\quad{\rm and}\quad
h\left\langle\phi\right\rangle\psi_{\ast} < M_{P}\Gamma\,.
\label{Eq.: Psi Energy Density Oscillating Constraint b}
\end{equation}
Substituting $\left\langle\phi\right\rangle$, $\Gamma$ and $\psi_{\ast}$, 
\cref{Eq.: Phi VEV,Eq.: Psi_{*}} and using that $\Gamma\sim g^2m$ (with $g<1$)
into \cref{Eq.: Psi Energy Density Oscillating Constraint b} gives the 
constraints
\begin{align}
h &> 10^{-4}g^{-2}\lambda^{-1/4}\,
\frac{M_P}{H_*}\left(\frac{m_{0}}{M_{P}}\right)^{3/2}
\quad{\rm and}\quad\nonumber\\
m_\psi &<(10^2gh)^2\frac{M_PH_*}{m_0}\,.
\label{Eq.: h Constraint d - End of Inflation Section}
\end{align}

\subsection{Results}
\label{Subsection: ``End of Inflation'' Mechanism Results}

We now combine the above constraints to find out the allowed parameter space.
%

\subsubsection{The parameter space}

From \cref{Eq.: g << 1 (Therm. Fluc. phi Subsubsection)} we require $g<1$.
We also require the constraint given by \cref{Eq.: Gamma_{varphi} Constraint a} 
to be satisfied, where $g$ is present as $g^{-16}$. Therefore, this latter 
constraint will start to become very strong very quickly as we decrease $g$. We
find that a value of $g\!=\!0.4$ yields allowed parameter space, for reasonable
values of $H_{\ast}$ and $\Gamma_{\varphi}$. The parameter space that we find here
however, when all constraints are considered together and regardless of the 
$\phi_{\ast}$ case, is actually a {\em sharp prediction} of single values for 
all but one of the free parameters and the other quantities in the model, to 
within an order of magnitude, rather than a range of parameter space. The 
values of the free parameters are displayed in 
\cref{Table: Free Parameter Values (alpha=1 and Gamma_{varphi}<<H_{TI})}.\\
\begin{table}[h!]
\centering
\begin{tabular}{c @{\hskip 30pt} c}
Parameter & Value\\
\hline\\
[5pt]
$g$ & $0.4$\\
[5pt]
$H_{\ast}$ & $10^{8}\ \mbox{GeV}$\\
[5pt]
$\Gamma_{\varphi}$ & $
10^{-6}\ \mbox{GeV}$\\
[5pt]
$\lambda$ & $
10^{-11}$\\
[5pt]
$h$ & $10^{-9}$
\end{tabular}
\caption{Values of the free parameters for which parameter space exists} 
\label{Table: Free Parameter Values (alpha=1 and Gamma_{varphi}<<H_{TI})}
\end{table}


Within the range $m_{0}\!\sim\!10^{2}\,\text{--}\,10^{3}\ \mbox{GeV}$,
the mass $m_{\psi}$ can span many orders of magnitude, with only an upper limit 
of $\sim\!10^{-4}\,\text{--}\,10^{-2}\ \mbox{GeV}$. Within the model, there is no
effective lower bound on $m_{\psi}$, but, of course, this cannot decrease too 
much.\footnote{Note that $\psi$ is much more massive today as its mass receives
a contribution due to the coupling with $\langle\phi\rangle$.} 
%
%
%
Values of other quantities in the model for a mass value of 
$m_{0}\!\sim\!10^{3}\ \mbox{GeV}$ and the parameter values of 
\cref{Table: Free Parameter Values (alpha=1 and Gamma_{varphi}<<H_{TI})} are 
shown in \cref{Table: Quantity Values for m_{0}sim10^{3} 
(alpha=1 and Gamma_{varphi}<<H_{TI})}. In this table we include the tensor 
fraction, which for $H_{\ast}\!\sim\!10^{8}\ \mbox{GeV}$ yields the 
negligible value $r\!\sim\!10^{-13}$.\\
\begin{table}[h!]
\centering
\begin{tabular}{c @{\hskip 30pt} c}
Quantity & 
Value\\
\hline
\\
[-8pt]
$\psi_{\ast}$ & $10^{12}\ \mbox{GeV}$\\
[5pt]
$\delta\psi_{\ast}/\psi_{\ast}$ & $10^{-4}$\\
[5pt]
$H_{TI}$ & $10^{-2}\ \mbox{GeV}$\\
[5pt]
$\left\langle\phi\right\rangle\sim\phi_{\rm BD}$ & $10^{13}\ \mbox{GeV}$\\
[5pt]
$V_{0}^{1/4}$ & $10^{8}\ \mbox{GeV}$\\
[5pt]
$T_{1}$ & $10^{7}\ \mbox{GeV}$\\
[5pt]
$T_{2}$ & $10^{3}\ \mbox{GeV}$\\
[5pt]
$\Gamma$ & $10^{2}\ \mbox{GeV}$\\
[5pt]
$r$ & $10^{-13}$
\end{tabular}
\caption{Values of quantities in the model for 
$m_{0}\!\sim\!10^{3}\ \mbox{GeV}$ and the parameter values of 
\cref{Table: Free Parameter Values (alpha=1 and Gamma_{varphi}<<H_{TI})}.}
\label{Table: Quantity Values for m_{0}sim10^{3} 
(alpha=1 and Gamma_{varphi}<<H_{TI})}
\end{table}

\subsubsection{\boldmath 
Values of $n_{s}$ and $n_{s}^{\prime}$ with 
quadratic chaotic inflation}

We provide results for the spectral index and its running when the period of 
primordial inflation is that of slow-roll quadratic chaotic inflation, with the
potential
\begin{equation}
V(\varphi) = \frac{1}{2}m_{\varphi}^{2}\varphi^{2}\,.
\label{Eq.: Chaotic Inflation Potential}
\end{equation}
From \cref{Subsection: Spectral Index --- n_{s} and n_{s}^{'}}, the spectral 
index $n_{s}$ is given by
\begin{equation}
n_{s} \simeq 1 - 2\epsilon + 2\eta_{\psi}\;,
\label{Eq.: Spectral Index - End of Inflation Results Section}
\end{equation}
with $\epsilon$ and $\eta_{\psi}$ being given by 
\cref{Eq.: epsilon and eta_{psi psi} Definition} 
and where both are to be evaluated at the point where cosmological scales exit 
the horizon during primordial inflation. The potential of 
\cref{Eq.: Chaotic Inflation Potential} gives
\begin{equation}
\epsilon = \frac{2M_{P}^{2}}{\varphi_{\ast}^{2}}\,.
\label{Eq.: epsilon - varphi_{*}}
\end{equation}
We obtain an expression for $\varphi_{\ast}$ in terms of $N_{\ast}$ by using the 
equation
\begin{equation}
N_{\ast}\simeq\frac{1}{M_{P}^{2}}\int_{\varphi_{\rm end}}^{\varphi_{\ast}}
\frac{V(\varphi)}{V^{\prime}(\varphi)}\,\mathrm{d}\varphi\,.
\label{Eq.: N_{*}}
\end{equation}
We define the end of primordial inflation to be when $\epsilon\!=\!1$. This 
gives \mbox{$\varphi_{\rm end} = \sqrt{2}M_{P}\,$}. Therefore we have
\begin{equation}
\varphi_{\ast} \simeq \sqrt{4N_{\ast}+2}\,M_{P}\;.
\label{Eq.: varphi_{*}}
\end{equation}
Substituting \cref{Eq.: varphi_{*}} into \cref{Eq.: epsilon - varphi_{*}} gives
\begin{equation}
\epsilon \simeq \frac{1}{2N_{\ast}+1}\,.
\label{Eq.: epsilon}
\end{equation}
We also need to calculate $\eta_{\psi}$. Using our potential, 
\cref{Eq.: Full Potential}, at the time cosmological scales exit the horizon, 
we obtain
\begin{equation}
\left.\frac{\partial^2 V}{\partial\psi^2}\right|_{\ast} = m_{\psi}^{2} + 
2h^{2}\phi_{\ast}^{2}\,.
\label{Eq.: V_{psi psi}|_{*}}
\end{equation}
Therefore we obtain $\eta_{\psi}$ as
\begin{equation}
\eta_{\psi} = \frac{1}{3H_{\ast}^{2}}\left(m_{\psi}^{2} + 
2h^{2}\phi_{\ast}^{2}\right).
\label{Eq.: eta_{psi psi}}
\end{equation}
Our final result for the spectral index is therefore
\begin{equation}
n_{s} \simeq 1 - \frac{2}{2N_{\ast}+1} + \frac23
\frac{m_{\psi}^{2} + 2h^{2}\phi_{\ast}^{2}}{H_{\ast}^{2}}\,.
\label{Eq.: n_{s}}
\end{equation}

From \cref{Subsection: Spectral Index --- n_{s} and n_{s}^{'}}, the running of 
the spectral index $n_{s}^{\prime}$ is given by
\begin{equation}
n_{s}^{\prime} \simeq -8\epsilon^{2} + 4\epsilon\eta + 4\epsilon\eta_{\psi}\;,
\label{Eq.: Running of Spectral Index - End of Inflation Results Section}
\end{equation}
with $\eta$ being given by \cref{Eq.: eta Definition}, which is to be evaluated 
at the point where cosmological scales exit the horizon during primordial 
inflation. The potential of \cref{Eq.: Chaotic Inflation Potential} gives
\mbox{$\eta=\epsilon$}, given by \cref{Eq.: epsilon - varphi_{*}}. Thus,
\begin{equation}
\eta \simeq \frac{1}{2N_{\ast}+1}\,.
\label{Eq.: eta}
\end{equation}

Our final result for the running of the spectral index is therefore
\begin{equation}
n_{s}^{\prime} \simeq -\frac{4}{\left(2N_{\ast}+1\right)^{2}} + 
\frac{4}{6N_{\ast}+3}\,
\frac{m_{\psi}^{2} + 2h^{2}\phi_{\ast}^{2}}{H_{\ast}^{2}}\,.
\label{Eq.: n_{s}'}
\end{equation}
Using the values in \cref{Table: Free Parameter Values 
(alpha=1 and Gamma_{varphi}<<H_{TI}),Table: Quantity Values for m_{0}sim10^{3} 
(alpha=1 and Gamma_{varphi}<<H_{TI})}, it is straightforward to show
that, for \mbox{$m_0\sim 10^3\,$GeV}, we have 
\mbox{$3\eta_\psi=\left(m_{\psi}^{2}+2h^{2}\phi_{\ast}^{2}\right)/H_*^2\sim 10^{-8}$}.
Therefore, the last term on the right-hand-side of 
\cref{Eq.: n_{s},Eq.: n_{s}'} is negligible.

In order to obtain $n_{s}$ and $n_{s}^{\prime}$, we first need to obtain $N_{\ast}$.
The values of $N_{\textrm{TI}}$ and $N_{\ast}$ in our the model are shown in 
\cref{Figure: Ns} respectively, with $g$, $H_{\ast}$, 
$\Gamma_{\varphi}$ and $\lambda$ values from 
\cref{Table: Free Parameter Values (alpha=1 and Gamma_{varphi}<<H_{TI})}. 
The kink that is visible in the plot of $N_{\ast}$ at around 
$m_{0}\!\sim\!10^{9}\ \mbox{GeV}$ is a result of the fact that for $m_{0}$ 
values larger than this, we do not have any period of thermal inflation, as can 
be seen in the plot of $N_{\textrm{TI}}$. The values of $N_{\textrm{TI}}$ and 
$N_{\ast}$ for a thermal waterfall field mass of $m_{0}\!\sim\!10^{3}\ \mbox{GeV}$
are shown in 
\cref{Table: N_{TI} and N_{*} for m_{0}sim10^{3} (Gamma_{varphi}<<H_{TI})}.
\begin{figure}[h!]
\begin{center}
\includegraphics[scale=.35]{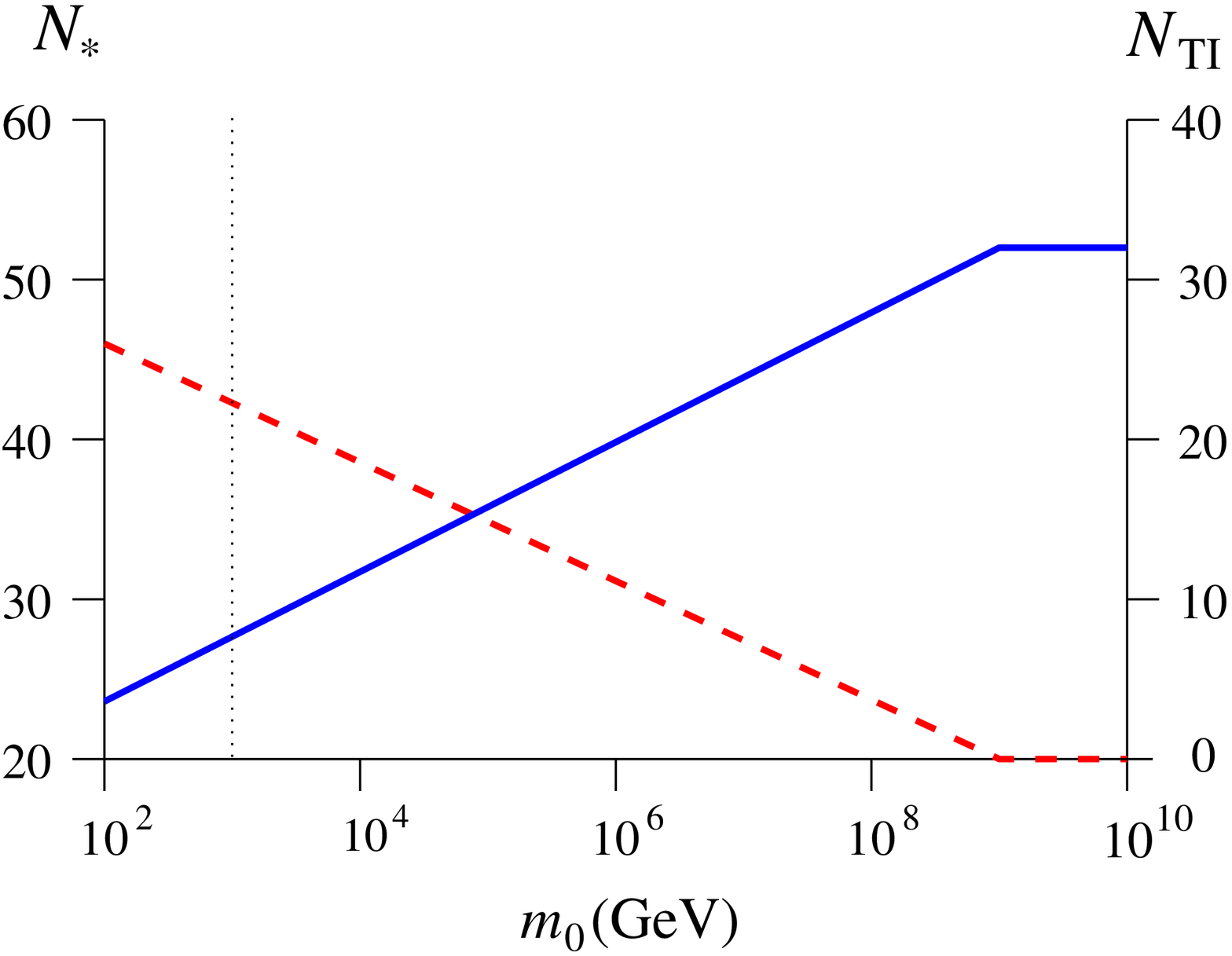}
\caption{Values of $N_*$ and $N_{\textrm{TI}}$ in our model, 
with $\Gamma_{\varphi}\!\ll\!H_{TI}$ and $g$, $\Gamma_{\varphi}$ and
$\lambda$ values from 
\cref{Table: Free Parameter Values (alpha=1 and Gamma_{varphi}<<H_{TI})}. 
(Plots of \cref{Eq.: N (Gamma_{varphi} << H_{TI}),%
Eq.: N_{*} (Gamma_{varphi} << H_{TI})}, with $m\!=\!m_{0}$.)
The \textcolor{blue}{Blue} solid line depicts $N_*$ and the 
\textcolor{red}{Red} dashed line depicts $N_{\rm TI}$,
such that $N_*+N_{\rm TI}=52$. The vertical dotted line depicts values for
\mbox{$m_0=10^3\,$GeV}.}
\label{Figure: Ns}
\end{center}
\end{figure}
%
%
\begin{table}[h!]
\centering
\begin{tabular}
{c @{\hskip 30pt} c}
Parameter & Value\\
\hline
\\
[-8pt]
$N_{\textrm{TI}}$ & $24$\\
[5pt]
$N_{\ast}$ & $28$
\end{tabular}
\caption{Values of $N_{\textrm{TI}}$ and $N_{\ast}$ in our model, with 
$\Gamma_{\varphi}\!\ll\!H_{\textrm{TI}}$, $m_{0}\!\sim\!10^{3}\ \mbox{GeV}$ and $g$, 
$H_{\ast}$, $\Gamma_{\varphi}$ and $\lambda$ values from 
\cref{Table: Free Parameter Values (alpha=1 and Gamma_{varphi}<<H_{TI})}.}
\label{Table: N_{TI} and N_{*} for m_{0}sim10^{3} (Gamma_{varphi}<<H_{TI})}
\end{table}
\\

The predicted values of $n_{s}$ and $n_{s}^{\prime}$ of the model for a thermal 
waterfall field mass of $m_{0}\!\sim\!10^{3}\ \mbox{GeV}$ in all cases of
$\phi_{\ast}$ are the same to within at least four significant figures. 
They are also both insensitive to the value of $m_{\psi}$ within its allowed 
range. $n_{s}$ and $n_{s}^{\prime}$ are shown in 
\cref{Table: n_{s} and n_{s}' for m_{0}sim10^{3} 
(alpha=1 and Gamma_{varphi}<<H_{TI})}, with them both being within current 
observational bounds \cite{Planck_2015_Cosmo._Results}.
The prediction of the model for $n_{s}$ and 
$n_{s}^{\prime}$ and for a spectator field mass at the upper bound of 
$m_{\psi}\!=\!10^{-2}\ \mbox{GeV}$ are shown in 
\cref{Figure: n_{s} for phi_{*} Case A,Figure: n_{s}' for phi_{*} Case A}
with the parameter values of 
\cref{Table: Free Parameter Values (alpha=1 and Gamma_{varphi}<<H_{TI})}. 

\begin{figure}[h!]
\begin{center}
\includegraphics[scale=.8]{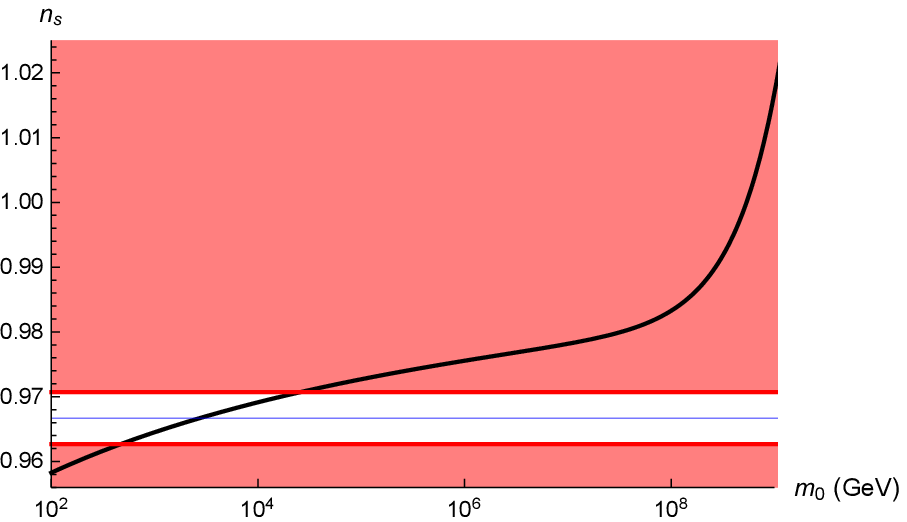}
\caption[Prediction for $n_{s}$ for $\phi_{\ast}$ Case A: Chaotic Inflation]%
{Prediction of the model for $n_{s}$ 
with primordial 
inflation being quadratic chaotic inflation $\Gamma_{\varphi}\!\ll\!H_{TI}$, 
$m_{\psi}\!=\!10^{-2}\ \mbox{GeV}$ and the parameter values from 
\cref{Table: Free Parameter Values (alpha=1 and Gamma_{varphi}<<H_{TI})}. 
(A plot of \cref{Eq.: n_{s}}, irrespective of the value of $\phi_{\ast}$, 
with $m\!=\!m_{0}$ and 
$\Gamma\!=\!g^{2}m_{0}$.) The \textcolor{blue}{Blue} and \textcolor{red}{Red} 
lines are the central value and lower/upper bounds of $n_{s}$, respectively, as 
obtained by the Planck mission \cite{Planck_2015_Cosmo._Results}.}
\label{Figure: n_{s} for phi_{*} Case A}
\end{center}
\end{figure}
\begin{figure}[h!]
\begin{center}
\includegraphics[scale=.8]{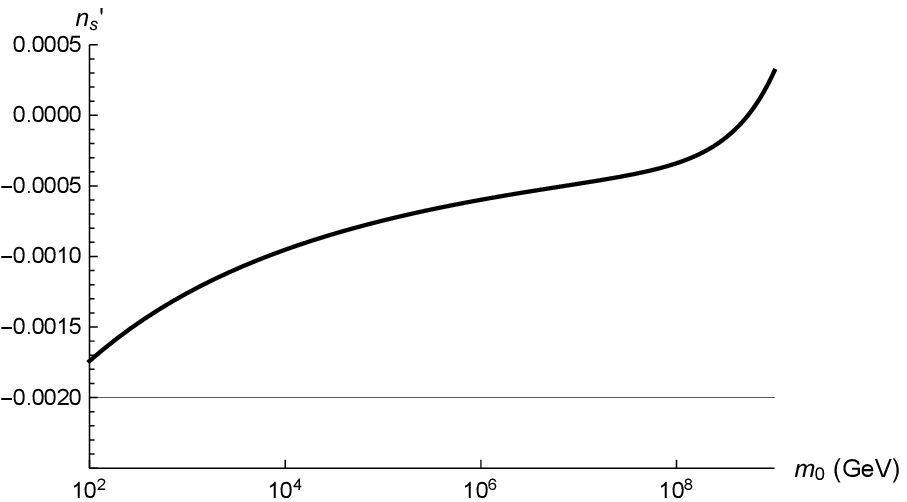}
\caption[Prediction 
for $n_{s}^{\prime}$ for $\phi_{\ast}$ Case A: Chaotic Inflation]%
{Prediction of the model for $n_{s}^{\prime}$ 
with primordial inflation being quadratic chaotic inflation, 
with $\Gamma_{\varphi}\!\ll\!H_{TI}$, $m_{\psi}\!=\!10^{-2}\ \mbox{GeV}$ and the 
parameter values from 
\cref{Table: Free Parameter Values (alpha=1 and Gamma_{varphi}<<H_{TI})}. 
(A plot of \cref{Eq.: n_{s}'}, irrespective of the value of $\phi_{\ast}$, 
with $m\!=\!m_{0}$ and 
$\Gamma\!=\!g^{2}m_{0}$.) The \textcolor{blue}{Blue} line is the central value 
of $n_{s}^{\prime}$ as obtained by the Planck mission 
\cite{Planck_2015_Cosmo._Results}, with the lower and upper bounds being outside the displayed range of $n_{s}^{\prime}$.}
\label{Figure: n_{s}' for phi_{*} Case A}
\end{center}
\end{figure}

\begin{table}[h!]
\centering
\begin{tabular}
{c @{\hskip 30pt} c}
Quantity & Value\\
\hline
\\
[-8pt]
$n_{s}$ & $0.9645$\\
[5pt]
$n_{s}^{\prime}$ & $-0.001259$
\end{tabular}
\caption{Prediction for $n_{s}$ and $n_{s}^{\prime}$ of the model with primordial 
inflation being quadratic chaotic inflation, with 
$\Gamma_{\varphi}\!\ll\!H_{\textrm{TI}}$, 
$m_{\psi}\!=\!10^{-2}\ \mbox{GeV}$, $m_{0}\!\sim\!10^{3}\ \mbox{GeV}$ and the 
parameter values from 
\cref{Table: Free Parameter Values (alpha=1 and Gamma_{varphi}<<H_{TI})}.}
\label{Table: n_{s} and n_{s}' for m_{0}sim10^{3} 
(alpha=1 and Gamma_{varphi}<<H_{TI})}
\end{table}

\section{Conclusions}
\label{Section: Conclusions}

We have thoroughly investigated a new model of thermal inflation, where the 
thermal waterfall field is coupled to a spectator field, which is
responsible for the observed primordial curvature perturbation through the 
``end of inflation'' mechanism. We have derived a multitude of 
constraints for the model parameters. We have found that the allowed parameter 
space for our model corresponds to a sharp prediction for inflationary 
observables, like the spectral index and its running. Taking quadratic chaotic 
inflation as an example, we have obtained the values shown in 
\cref{Table: n_{s} and n_{s}' for m_{0}sim10^{3} 
(alpha=1 and Gamma_{varphi}<<H_{TI})},
which are in excellent agreement with the latest Planck data (well within 
1-$\sigma$). We also found negligible tensors, with $r\sim 10^{-13}$.

Our model works with tachyonic mass for our thermal waterfall field that is of 
order 1~TeV. This is rather natural for a flaton field, which corresponds to
a flat direction in supersymmetry lifted by a soft mass 
\cite{Lyth_&_Stewart:Therm._Inf._Moduli_Prob.,
Barreiro_et_al.:Aspects_Therm._Inf.:_Finite_Temp._Pot._Top._Defects,
Asaka_&_Kawasaki:Cos._Moduli_Problem_Therm._Inf._Models}. The energy scale
of primordial and of thermal inflation were found to be $10^{13}\,$GeV and
$10^8\,$GeV respectively, which are very reasonable values. Notice that 
low-scale primordial inflation ensures that the contribution to the curvature 
perturbation of the inflaton field is negligible. 

It should be stressed that the choice of model for primordial inflation may 
differ from our quadratic chaotic inflation example. We have found that, in the
allowed parameter space, the direct contribution of our spectator field to $n_s$
and $n_s'$ is negligible as $\eta_\psi\sim 10^{-8}$. Thus, our expressions in 
\cref{Eq.: Spectral Index - Spectral Index Section,%
Eq.: Running of Spectral Index - Spectral Index Section} become
\mbox{$n_s\simeq 1-2\epsilon$} and 
\mbox{$n_s'\simeq 8\epsilon^2+4\epsilon\eta$}. Therefore, given a particular 
model of primordial inflation, it is straightforward to evaluate the slow-roll 
parameters $\epsilon$ and $\eta$ and find $n_s$ and $n_s'$. 

The number $N_*$ of remaining e-folds of primordial inflation when the 
cosmological scales exit the horizon is drastically reduced by the presence of 
a subsequent period of thermal inflation. In the allowed parameter space, 
\mbox{$N_*\simeq 28$}. This determines the values of $\epsilon$ and $\eta$ and
in turn the observables $n_s$ and $n_s'$. Note that our $N_*$ is substantially 
smaller than the usual 60 e-folds. Consequently, the produced values of 
$n_s$ and $n_s'$ may vary substantially from the usual numbers corresponding to 
the particular model of primordial inflation considered. This can render viable 
inflationary models that would be otherwise excluded by observations.\footnote{%
Such reconciliation of high scale models of inflation may also occur using 
non-standard initial conditions for fluctuations \cite{recon}.} This 
effect of a period of thermal inflation resurrecting inflationary models has 
been employed in Ref.~\cite{charlotte}. 

Note also that, in our case, thermal inflation can last much longer that the 
typical 10-15 e-folds, because we have considered that reheating
for primordial inflation occurs after thermal inflation. So, the above effect, 
i.e. modifying the inflationary observables by changing $N_*$ due to thermal 
inflation, is intensified.

All in all, we have thoroughly investigated a new model of thermal inflation, 
in which the curvature perturbation is due to a spectator field coupled to the 
thermal waterfall field. For natural values of the model's mass scales,
we have found a sharp prediction of inflationary observables that depends on the
chosen model of primordial inflation. Considering quadratic chaotic inflation 
resulted in numbers that are in excellent agreement with Planck observations.
Our paper serves to remind readers that realistic models of inflation, in which 
the curvature perturbation is not generated by the inflaton field, are viable 
alternatives to the simple single-field inflation paradigm. 

\section*{Acknowledgements}
AR thanks Anupam Mazumdar for several helpful discussions regarding 
thermalization and thermal interaction rates. The work of KD and DHL is 
supported by the Lancaster-Manchester-Sheffield Consortium for Fundamental 
Physics under STFC grant ST/L000520/1. 
The early part of the work of AR was funded by an STFC PhD studentship.

\addcontentsline{toc}{section}{References}
\providecommand{\href}[2]{#2}
\begingroup\raggedright

\endgroup

\end{twocolumn}

\end{document}